\documentclass[10pt]{revtex4}
\usepackage{amsmath}
\usepackage{hyperref}

\begin{document}

\title{{\Large Comments on interactions in the SUSY models}}
\author{ Sudhaker Upadhyay${}^{a}$\thanks{%
e-mail address: sudhakerupadhyay@gmail.com}}
\author{ Alexander Reshetnyak${}^{b}$\thanks{%
e-mail address: reshet@ispms.tsc.ru}}
\author{ Bhabani Prasad Mandal${}^{a}$\thanks{%
e-mail address: bhabani.mandal@gmail.com}}
\affiliation{${}^{a}$Department of Physics, Banaras Hindu University, Varanasi-221005,
India \\
${}^{b}$Institute of Strength Physics and Materials Science of SB RAS, 634021, Tomsk,
Russia}

\begin{abstract}
We consider special supersymmetry (SUSY) transformations with $m$ generators
$\overleftarrow{s}_{\alpha }$ for a certain class of models and study some
physical consequences of Grassmann-odd transformations which form an Abelian
supergroup with finite parameters and respective group-like elements being
functionals of field variables. The SUSY-invariant path integral measure
within conventional quantization implies the appearance, under a change of
variables related to such SUSY transformations, of a Jacobian which is
explicitly calculated. The Jacobian implies, first of all, the appearance of
trivial interactions in the transformed action, and, second, the presence of
a modified Ward identity which reduces to the standard Ward identities in the
case of constant parameters. We examine the case of ${N}=1$ and $N=2$
supersymmetric harmonic oscillators to illustrate the general concept by a
simple free model with $(1,1)$ physical degrees of freedom. It is shown that
the interaction terms $U_{tr}$ have a corresponding SUSY-exact form: $U_{tr}=%
\big(V_{(1)}\overleftarrow{s};V_{(2)}\overleftarrow{\bar{s}}\overleftarrow{s}%
\big)$ naturally generated in this generalized formulation. We argue that
the case of non-trivial interactions cannot be obtained in such a way.
\end{abstract}

\maketitle



\section{Introduction}

Supersymmetric theories are invariant under SUSY transformations which
relate the bosonic and fermionic degrees of freedom present in the theories
and were initially proposed with the motivation of studying the fundamental
interactions in a unified manner. The generators of SUSY transformations
satisfy Lie superalgebra relations which are closed under the combination of
commutators and anticommutators. Local or nonlinear versions of the Lie
superalgebra construction were extended to various field-theoretic models,
such as superstring theories \cite{superstring}, supergravity \cite{fer, des}
(for modern developments, see Refs. \cite{buchbinder, ivanov}) and
higher-spin field theories \cite{reviewsV, reviews3, symferm-ads,
Reshetnyk2, MetsaevBV1, Zinoviev}. SUSY theories provide a bosonic
superpartner to each fermion present in a theory and vice-versa. This
indicates that if $N=1$ SUSY\ (with one fermionic generator in terms of Dirac spinor) is to be a
perfect symmetry of nature, then each set of superpartners must have the
same set of quantum numbers with the only difference in spin. Despite the
beauty of all such unified theories, SUSY has not been supported by
experimental evidence so far, but remains one of the problems included in
LHC experimental program.

Some variants of SUSY have also provided an interesting topic in quantum
mechanics \cite{avi} due to a link to exactly solvable models. SUSY and its
breaking have been studied in various simple quantum mechanical systems
involving a spin-1/2 particle moving in one direction \cite{wit,coo}. The
supersymmetric Hamiltonian may be presented in terms of supercharges which
generate SUSY transformations. A path integral formulation of SUSY in
quantum mechanics was first analysed by Salomonson and van Holten \cite{sal}%
. Further, by using SUSY methods, the rate of tunnelling through quantum
mechanical barriers was accurately determined \cite{31, 32, 33, 34}.

SUSY transformations, when applied to gauge theory, together with special
global SUSY transformations known as BRST transformations \cite{brs,brs1},
have also been explored in a more effective way \cite{rup,van}. The BRST
symmetry and the associated concept of BRST cohomology provide the commonly
used quantization methods in Lagrangian \cite{deWittH, bv} and Hamiltonian
\cite{bfv, bfv1} formalisms for gauge and string theories \cite{ht,wei}. The
BRST symmetry was generalized \cite{bv}\ to the case of an infinitesimal
field-dependent (FD) transformation parameter $\mu $, $\mu ^{2}=0$, within
the field-antifield formalism \cite{deWittH, bv}  in order to prove the
independence from small gauge variations of the path integral for arbitrary
gauge theories. A further generalization \cite{sdj} was made in Yang--Mills
theories with $R_{\xi }$-gauges by making the transformation parameter
finite and field-dependent, as one considers a sequence of infinitesimal
field-dependent BRST transformations (for a numeric parameter $\kappa $)
with some applications \cite{upa, sdj1,rb,susk,sb,
bss,smm,fs,sud1,sudhak,rbs,das,rs,su}. Another way to consider a finite
field-dependent parameter in Yang--Mills theories was inspired by a research
devoted to the so-called \emph{soft BRST symmetry breaking} problem \cite%
{llr1}, with reference to the Gribov problem \cite{Gribov}, which involves
the Zwanziger proposal \cite{Zwanziger} for a horizon functional joined
additively to a BRST-invariant quantum action. In fact, the horizon
functional in $R_{\xi }$-gauges with small $\xi $ was found explicitly \cite%
{llr1} (see Eq.~(5.20) therein) by using field-dependent BRST
transformations with a small odd-valued parameter, which was then extended
to be finite \cite{ll1}. The case of finite field-dependent BRST
transformations for general gauge theories was considered in \cite{re},
whereas for BRST-antiBRST symmetry \cite{aBRST1,aBRST2,aBRST3} in \cite{re1,
re5}, with the original calculation algorithm for functional Jacobians (for
a comparative analysis of BRST symmetry, see \cite{mr}).

At the same time, analogous properties of space-time SUSY transformations
(with Grassman-odd parameters) have never been generalized. Therefore, in
spite of the fact that BRST transformations are realized in an extended
field space with the initial classical, as well as the ghost, antighost and
Nakanishi--Lautrup fields and are reminiscent of gauge transformations, a
similar application of SUSY transformations in the path integral with FD
Grassman-odd parameters to field-theoretical models (without auxiliary field
variables introduced via the Faddeev--Popov prescription \cite{fp}) provides
us with an opportunity to apply the above research to the study of an
influence of SUSY transformations on the quantum action structure.

In this paper, we consider a generalization of SUSY transformations to the
case of $m$-parametric Lie superalgebra with the transformation parameters
being finite and field-dependent. In this way, the resulting transformations
remain a symmetry of the supersymmetric action. Under generalized SUSY
transformations with arbitrary field-dependent parameters the functional
measure, however, is not invariant. This leads to a non-trivial Jacobian for
the functional measure and therefore to a modification of the quantum action
by non-quadratic terms being a SUSY-exact contribution. For some choices of
parameters, the generalized SUSY transformations amount a precise change in
the exponent action. We illustrate these results by the example of a free
toy model with $(1,1)$ physical degrees of freedom, describing a
supersymmetric harmonic oscillator with the generalized $N=1$ and $N=2$ SUSY
transformations. In such a theory, the interaction terms emerge naturally in
the functional integral under thus generalized SUSY with specific parameters.

The paper is organized as follows. In Section~\ref{GenSUSY}, we study the
generalized SUSY transformations with $m$ Grassman-odd parameters for a
general supersymmetric invariant theory, calculate the corresponding
Jacobian of the change of variables, derive the standard and modified Ward
identities and classify the interactions. In Section~\ref{SUSYharm}, we
illustrate the example of generalized SUSY transformations by a
supersymmetric harmonic oscillator with $(1,1)$ degrees of freedom, in such
a way that the trivial interaction terms are produced by generalized $N=1$, $%
N=2$ SUSY transformations from the functional measure. Finally, we summarize
the results in Conclusions.

We use the DeWitt condensed notation and the conventions of \cite{sdj, re,
re1}, e.g., $\varepsilon (F)$ denotes the value of Grassmann parity of a
quantity $F$.

\vspace{-1ex}

\section{Generalized SUSY transformations}

\label{GenSUSY} Here, we investigate a finite field-dependent SUSY (FSUSY)
transformation for general supersymmetric invariant theories (following the
techniques developed in both \cite{sdj} and \cite{re, re1}). To this end, we
first define a SUSY transformation with infinitesimal Grassmann-odd constant
parameters $\epsilon ^{\alpha }$, $\alpha =1,..,m$, $\varepsilon (\epsilon
^{\alpha })=1$, leaving invariant an action $S(q)$ of generic variables $%
q^{i}$, $i=1,...,n$, $n=(n_{+},n_{-})$, $\varepsilon (q^{i})=\varepsilon
_{i} $:
\begin{equation}
\delta _{\epsilon }q^{i}=\mathcal{R}_{\alpha }^{i}(q)\epsilon ^{\alpha
}=q^{i}\overleftarrow{s}_{\alpha }\epsilon ^{\alpha }:\quad S(q+\delta
_{\epsilon }q)=S(q)+o(\epsilon)\Longleftrightarrow S(q)\overleftarrow{\partial }_{i}%
\mathcal{R}_{\alpha }^{i}(q)=S(q)\overleftarrow{s}_{\alpha }=0,  \label{SUSY}
\end{equation}%
where $\overleftarrow{\partial }_{i}\equiv \frac{\overleftarrow{\delta }}{%
\delta q^{i}}$ and $\mathcal{R}_{\alpha }^{i}(q)$, $\overleftarrow{s}%
_{\alpha }$, $\varepsilon \big(\mathcal{R}_{\alpha }^{i},\overleftarrow{s}%
_{\alpha }\big)=\big(\varepsilon _{i}+1,1\big)$ are, respectively,
the generators of SUSY transformations acting on the variables $q^{i}$%
and those acting on functionals $F(q)$. We suppose that the
generators of SUSY transformations satisfy Abelian anticommutator relations:
\begin{equation}
\lbrack \delta _{\epsilon _{1}},\delta _{\epsilon
_{2}}]=0\Longleftrightarrow \{\overleftarrow{s}_{\alpha },\overleftarrow{s}%
_{\beta }\}=0.  \label{SUSYalg}
\end{equation}%
The group transformations with finite parameters, $\epsilon ^{\alpha }$, $%
q^{i}\rightarrow q^{\prime i}=q^{i}(q|\epsilon )$, may be restored by two
equivalent ways from the Lie equations and from the requirement for any $%
\overleftarrow{s}_{\alpha }$ -closed functional $F(q)$ to be invariant with
respect to right group transformations:
\begin{equation}
q^{i}(q|\epsilon )\overleftarrow{\partial }_{\alpha }=q^{i}(q|\epsilon )%
\overleftarrow{s}_{\alpha }\ \left( \mathrm{for}\ \overleftarrow{\partial }%
_{\alpha }\equiv \frac{\overleftarrow{\partial }}{\partial \epsilon ^{\alpha
}}\right) \Longleftrightarrow F\big(q^{i}(q|\epsilon )\big)=F(q).
\label{Lieeq}
\end{equation}%
For a $t$-rescaled argument $\epsilon ^{\alpha }\rightarrow t\epsilon
^{\alpha }$ of $q^{i}(q|t\epsilon )$, the form of Lie equations is
equivalent to (\ref{Lieeq}) with a formal solution for constant $\epsilon
^{\alpha }$%
\begin{equation}
\frac{d}{dt}q^{i}(q|t\epsilon )=q^{i}(q|t\epsilon )\overleftarrow{s}_{\alpha
}\epsilon ^{\alpha }\Longrightarrow q^{i}(q;t\epsilon )= q^{i}\exp \{t%
\overleftarrow{s}_{\alpha }\epsilon ^{\alpha }\},  \label{Liemod}
\end{equation}%
so that the set of finite transformations forms an Abelian group $%
G=\{g(\epsilon ):g(\epsilon )=\exp \{t\overleftarrow{s}_{\alpha }\epsilon
^{\alpha }\},\ g(\epsilon _{1})g(\epsilon _{2})=g(\epsilon _{2})g(\epsilon
_{1})\}$, for constant $\epsilon ^{\alpha }$. For field-dependent $\epsilon
^{\alpha }=\epsilon ^{\alpha }(q)$ having no explicit dependence on
space-time coordinates $x^{\mu }$, $\partial _{\mu }\epsilon ^{\alpha }(q)=0$%
, the set of algebraic elements $\mathcal{G}=\{\tilde{g}_{lin}(\epsilon
(q)):=1+\overleftarrow{s}_{\alpha }\epsilon ^{\alpha }(q)\}$ forms a
non-linear algebra which corresponds to a set of formal group-like finite
elements:
\begin{equation}
\tilde{G}=\left\{ \tilde{g}(\epsilon (q)):\tilde{g}=1+\sum_{i=1}^{m}\frac{1}{%
i!}\Big[\prod_{k=1}^{i}(\overleftarrow{s}_{\alpha
_{k}})\prod_{k=1}^{i}\epsilon ^{\alpha _{k+1-i}}(q)\Big]\right\} ,
\label{tildeG}
\end{equation}%
obtained as solutions for $F\big(q^{i}(q|\epsilon (q))\big)=F(q)$ in (\ref%
{Lieeq}) with finite FD $\epsilon ^{\alpha }(q)$, as in \cite{re1}. Note
that in the case $m=1,2$ we have representations of finite BRST and
BRST-antiBRST group (\ref{Liemod}) and group-like elements (\ref{tildeG})
\cite{re, re1}.

We refer to SUSY transformations generalized in such a way as FSUSY
transformations. Another way to derive FSUSY transformations for an $N=1$%
-parametric subset from $G$, i.e., for $m=1$ can be done by rendering the
infinitesimal parameter $\epsilon ^{1}\equiv \epsilon $ field-dependent
through a continuous interpolation of an arbitrary parameter $\kappa \
(0\leq \kappa \leq 1)$, following Ref. \cite{sdj}: $(q^{i}(\kappa
=0);q^{i}(\kappa =1)=(q^{i};q^{i}(q|\epsilon )$.

An infinitesimal field-dependent SUSY transformation can be defined as
\begin{equation}
{dq^{i}(\kappa )}=\mathcal{R}^{i}(q)\epsilon ^{\prime }(q(\kappa ))%
{d\kappa },  \label{diff}
\end{equation}%
where $\epsilon ^{\prime }(q(\kappa )){d\kappa }$ is an infinitesimal
field-dependent parameter. An FSUSY transformation with a finite
field-dependent parameter can now be constructed by integrating such an
infinitesimal transformation from $\kappa =0$ to $\kappa =1$, as follows:%
\begin{equation}
q^{i}(q|\epsilon )\equiv q^{i}+\mathcal{R}^{i}(q)\epsilon (q),\ \mathrm{where%
}\ \epsilon (q)=\int_{0}^{1}d\kappa \epsilon ^{\prime }(q(q|\kappa )).
\label{kdep}
\end{equation}%
Note that in the case $m>1$ it is impossible to restore FSUSY
transformations in this way explicitly using (\ref{diff}) by a simple
integration over an auxiliary $\kappa $ (for details, see \cite{sdj}).

Turning to FSUSY transformations with $m$ odd parameters, we can see that
such transformations remain a symmetry of the supersymmetric action, which
is supposed to describe a non-degenerate non-gauge theory, whereas the path
integral measure will not be invariant under such transformations and
thereby will lead to a non-trivial Jacobian for a corresponding change of
variables in the generating functional $Z(J)$ of Green's functions with
external sources, $J_{i}$ ($\varepsilon (J_{i})=\varepsilon _{i}$) and in
the path integral $Z(0)=Z_{0}$:
\begin{eqnarray}
Z(J) &=&\int \mathcal{D}q\exp \left\{ \frac{\imath }{\hbar }\big[%
S(q)+J_{i}q^{i}\big]\right\} ,\ Z_{0}\ =\ \int \mathcal{I}_{q}\ \mathrm{with}%
\ \mathcal{I}_{q\tilde{g}(\epsilon )}\ =\ J(q)\mathcal{I}_{q},  \label{ZZ0}
\\
\ \mathrm{where} &&J(q)=\mathrm{Sdet}\left\Vert q^{i}(q|\epsilon (q))%
\overleftarrow{\partial }_{j}\right\Vert =\exp \hspace{-0.1em}\left\{
\hspace{-0.1em}\mathrm{Str}\,\mathrm{ln}\hspace{-0.1em}\left( \hspace{-0.1em}%
\delta _{j}^{i}+M_{j}^{i}(q,\epsilon )\right) \hspace{-0.1em}\right\} \
\mathrm{for}\ M_{j}^{i}(q,\epsilon )=\Delta q^{i}(q|\epsilon )\overleftarrow{%
\partial }_{j},  \label{sdet}
\end{eqnarray}%
which vanishes when $\epsilon ^{\alpha }=\mathrm{const}$, $%
M_{j}^{i}(q,\epsilon )\big|_{\epsilon =\mathrm{const}}=0$. The Jacobian can
be calculated explicitly, following the receipe \cite{re, re1, re5}, and
also by using the Green function method \cite{blt}. The latter approach,
using $t$-rescaled parameteres $t\epsilon ^{\alpha }$ (\ref{Liemod}) and the
inverse (formal) transformations $\tilde{g}{}^{-1}(\epsilon (q))$,
\begin{equation}
q^{i}(q|t\epsilon )\tilde{g}{}^{-1}(\epsilon (q))=q^{i}\Longrightarrow
q^{i}(q^{\prime }|t\epsilon )\overleftarrow{\partial }_{\alpha }=-tq^{i}%
\overleftarrow{s}_{\alpha },  \label{invLie}
\end{equation}%
assumes that the representation for $\ln J(q)$ given by (\ref{sdet}) reads
\begin{equation}
\ln J(q)=\mathrm{sTr}\ln \left( \hspace{-0.1em}\delta
_{j}^{i}-q^{i}(q^{\prime }|\epsilon )\overleftarrow{\partial }_{\alpha }\big(%
\epsilon ^{\alpha }\overleftarrow{%
\partial }_{j}\big)\right) \Longrightarrow
\frac{d}{dt}\ln J(q)=-\mathrm{tr}_{G}\left( [e+tm]^{-1}m\right) ,\ \
m_{\beta }^{\alpha }=\epsilon ^{\alpha }\overleftarrow{s}{}_{\beta },
\label{greenrepres}
\end{equation}%
where $\left( e\right) _{\beta }^{\alpha }$ and $\mathrm{tr}_{G}$ denote $%
\delta _{\beta }^{\alpha }$ and trace over matrix $G$ indices. In deriving (%
\ref{greenrepres}), we have used the fact that in differentiating with
respect to $t$, the first of the above equalities reads%
\begin{equation}
G_{j}^{i}[q^{j}\overleftarrow{s}_{\alpha }][\epsilon ^{\alpha }(q)%
\overleftarrow{\partial }_{i}](-1)^{\varepsilon _{i}}\ \mathrm{and}\ \mathrm{%
follows}\ \mathrm{from}\ G_{j}^{i}+t[q^{i}\overleftarrow{s}_{\alpha
}][\epsilon ^{\alpha }(q)\overleftarrow{\partial }_{k}]G_{j}^{k}=\delta
_{j}^{i}.  \label{greens}
\end{equation}%
From the latter representation, we find%
\begin{equation}
\epsilon ^{\alpha }(q)\overleftarrow{\partial }_{k}G_{j}^{k}=\left(
[e+tm]^{-1}\right) _{\beta }^{\alpha }\big(\epsilon ^{\beta }(q)%
\overleftarrow{\partial }_{j}\big),  \label{solGreen}
\end{equation}%
so that after substitution in the first term of (\ref{greens}) we get the
representation for the last quantity in (\ref{greenrepres}), which after
integration leads to the final result for the Jacobian (because $\ln
J(q(0))=0$)
\begin{equation}
J(q(\epsilon ))=\exp \Big\{-\mathrm{tr}_{G}\ln \left( [e+m]\right) \Big\}.
\label{jacobianres}
\end{equation}%
The Jacobian for $m=1,2$ is reduced to already known Jacobians for $N=1,2$
finite FD BRST transformations with nilpotent, $\overleftarrow{s},%
\overleftarrow{s}_{a}$, $a=1,2$. For functionally-independent FD $\epsilon
_{\alpha }(q)$ the Jacobian is not $\overleftarrow{s}_{\alpha }$-closed, in
general, whereas for $\overleftarrow{s}_{\alpha }$-potential (thereby
functionally-dependent) parameters
\begin{equation}
\hat{\epsilon}{}^{\alpha }(q)=\frac{1}{(m-1)!}\Lambda (q)\varepsilon
^{\alpha \alpha _{1}\ldots \alpha _{m-1}}\overleftarrow{s}_{\alpha
_{1}}\ldots \overleftarrow{s}_{\alpha _{m-1}},\ \mathrm{for}\ \varepsilon
^{12...m}=1\ \mathrm{and}\ \varepsilon ^{\alpha _{0}\alpha _{1}\ldots \alpha
_{m-1}}\varepsilon _{\alpha _{m-1}\ldots \alpha _{1}\alpha _{0}}={m!}
\label{fdepparam}
\end{equation}%
with an arbitrary potential functional, $\Lambda (q)$, $\varepsilon (\Lambda
)=m$, and totally antisymmetric tensors $\varepsilon ^{\alpha _{0}\alpha
_{1}\ldots \alpha _{m-1}}$, $\varepsilon _{\alpha _{m-1}\ldots \alpha
_{1}\alpha _{0}}$ the Jacobian is $\overleftarrow{s}_{\alpha }$-closed.

Due to the equivalence theorem \cite{tyutinkallosh}, the change of variables
in $Z(J)$ and in the path integral $Z_{0}$ generated by FSUSY
transformations (in terms of the integrand)
\begin{equation}
\mathcal{I}_{q\tilde{g}(\epsilon )}\ =\ J(q)\mathcal{I}_{q}=Dq\exp \left\{
\frac{\imath }{\hbar }\big[S(q)+\imath \hbar \mathrm{tr}_{G}\ln \left(
[e+m]\right) \big]\right\} =Dq\exp \left\{ \frac{\imath }{\hbar }\big[%
S(q)+S_{1}(q,\epsilon (q))\big]\right\} ,  \label{chanvar}
\end{equation}%
leads to the same quantum theory, $Z_{0}=Z_{\epsilon }$, with the same
conventional $S$-matrix. At the same time, a representation for the
transformed action, $S(q,\epsilon (q))=S(q)+S_{1}(q,\epsilon (q))$, should
be supersymmetrically invariant: $S(q,\epsilon (q))\overleftarrow{s}_{\alpha
}=0$. FSUSY transformations which satisfy the above must obey the condition
\begin{equation}
S_{1}(q,\epsilon (q))\overleftarrow{s}_{\alpha }=\imath \hbar \mathrm{tr}%
_{G}\ln \left( [e+m]\right) \overleftarrow{s}_{\alpha }=0,  \label{cond1}
\end{equation}%
In particular, $N=m$ FSUSY transformations $\tilde{g}(\hat{\epsilon}{}%
^{\alpha }(q))$ with FD parameters (\ref{fdepparam}) for any potential $%
\Lambda (q)$ satisfy the condition (\ref{cond1}).

Therefore, only\emph{\ trivial interactions} $U_{tr}(q)$ can be generated
(locally) by FSUSY transformations in the path integral, which are
characterized by the condition $U_{tr}(q)\overleftarrow{s}_{\alpha }=0$,
whereas the \emph{non-trivial interactions} $U(q)$ which lead to a different
$S$-matrix should satisfy the requirement
\begin{equation}
U(q)\overleftarrow{s}_{\alpha }=0:\quad U(q)\neq V^{\alpha }(q)%
\overleftarrow{s}_{\alpha }\forall V^{\alpha }(q).  \label{nontrivial}
\end{equation}%
For $m=1$, $m=2$ FSUSY transformations, the corresponding Jacobians (for
functionally dependent $\epsilon _{a}=\Lambda \overleftarrow{s}_{a}$, $%
(\epsilon _{a},\overleftarrow{s}^{a})$ = $(\varepsilon _{ab}\epsilon
^{b},\varepsilon ^{ab}\overleftarrow{s}_{b})$ with antisymmetric $%
\varepsilon _{ab}=-\varepsilon _{ba}$ and $\varepsilon ^{ab}$: $\varepsilon
^{ab}\varepsilon _{bc}=\delta _{c}^{a}$ , under the normalization $%
\varepsilon ^{12}=1$)
\begin{equation}
J_{(1)}(q(\epsilon ))=\left( 1+\epsilon \overleftarrow{s}\right)
^{-1}\Rightarrow J_{(1)}\overleftarrow{s}=0\ \mathrm{and}\ J_{(2)}\left(
q(\Lambda \overleftarrow{s}_{a})\right) =\left( 1+\frac{1}{2}\Lambda
\overleftarrow{s}_{a}\overleftarrow{s}^{a}\right) ^{-2}\Rightarrow J_{(2)}%
\overleftarrow{s}_{a}=0,  \label{jacobianres2}
\end{equation}%
lead only to trivial interactions. The invariance of the integrand $\mathcal{%
I}_{q}$ (\ref{ZZ0}) with respect to FSUSY with constant parameters $\epsilon
^{\alpha }$ leads to the presence of Ward identities for $Z(J)$:
\begin{equation}
J_{i}\langle q^{i}\overleftarrow{s}_{\alpha }\rangle _{J}=0\ \mathrm{where}\
\ \langle A(q)\rangle _{J}=Z^{-1}(J)\int \mathcal{D}qA(q)\exp \left\{ \frac{%
\imath }{\hbar }\big[S(q)+J_{i}q^{i}\big]\right\} ,\ \ \langle 1\rangle
_{J}=1,  \label{usWI}
\end{equation}%
with a source-dependent average expectation value for a certain functional $%
A(q)$ corresponding to a given action $S(q)$. In turn, the property (\ref%
{chanvar}), with account taken of (\ref{jacobianres}), for FD FSUSY
transformations means the presence of a so-called \emph{modified Ward
identity} depending on FD parameters $\epsilon (q)$:
\begin{equation}
\left\langle \exp \Big\{\frac{\imath }{\hbar }J_{i}q^{i}\sum_{i=1}^{m}\frac{1%
}{i!}\Big[\prod_{k=1}^{i}(\overleftarrow{s}_{\alpha
_{k}})\prod_{k=1}^{i}\epsilon ^{\alpha _{k+1-i}}(q)\Big]\Big\}\exp \Big\{-%
\mathrm{tr}_{G}\ln \left( [e+m]\right) \Big\}\right\rangle _{J}=1,
\label{modWI}
\end{equation}%
(for $m_{\beta }^{\alpha }=\epsilon ^{\alpha }(q)\overleftarrow{s}{}_{\beta
} $) which reduces to (\ref{usWI}) for constant $\epsilon ^{\alpha }$.

In the case $m=1$ (but not $m>1$) FSUSY may also be considered by evaluation
of the Jacobian according to \cite{sdj}, restricted by an infinitesimal FD
parameter $\epsilon ^{\prime }(q(\kappa ))$ according to the change of
variables $q^{i}(\kappa )\rightarrow q^{i}(\kappa +d\kappa )$ with Jacobian $%
J(\kappa )$ (\ref{sdet}):
\begin{eqnarray}
\mathcal{D}q(\kappa +d\kappa ) &=&J(\kappa )\mathcal{D}q(\kappa )\overset{def%
}{=}\mathcal{D}q(\kappa )\exp \left[ -(-1)^{i}\mathcal{R}^{i}(q)(\kappa )%
\big(\epsilon ^{\prime }(q(\kappa ))\overleftarrow{\partial }_{i}^{(q(\kappa
))}\big)\right] ,  \label{sdjdet} \\
&\mathrm{for}&J(\kappa )=1-\left[ (-1)^{i}\mathcal{R}^{i}(q)(\kappa )\big(%
\epsilon ^{\prime }(q(\kappa ))\overleftarrow{\partial }_{i}^{(q(\kappa ))}%
\big)\right] .
\end{eqnarray}%
As we suppose that after a change of variables generated by FSUSY
transformations $q^{i}\rightarrow q^{i}(\kappa )$ the supersymmetric action $%
S(q)$ also changes to $S(q)+S_{1}(\kappa )$ with a local functional $S_{1}(q)
$ vanishing at $\kappa =0$, the functional equation must hold%
\begin{equation}
\int \mathcal{D}q(\kappa )\left[ \frac{d}{d\kappa }\ln {J}(\kappa )-\frac{%
\imath }{\hbar }\frac{d}{d\kappa }S_{1}(q(\kappa ))\right] \exp \left\{
\frac{\imath }{\hbar }\big[S(q)+S_{1}(q(\kappa ))\big]\right\} .  \label{jac}
\end{equation}%
The necessary condition that equation (\ref{jac}) be solvable is $%
S_{1}(q(\kappa ))\overleftarrow{s}=0$, i.e., the addition to supersymmetric
action must also be supersymmetric. Once again, FSUSY transformations with
appropriate parameters $\epsilon $ change a supersymmetric action $S_{susy}$
to a new effective action $S_{susy}+S_{1}(\kappa =1)$ within functional
integration.

Note that one can perform a similar analysis in the case of $N=1$ SUSY
transformations with parameters $\bar{\epsilon}$ and the result will be the
same. The only difference is that the parameter $\epsilon $ will be replaced
everywhere by $\bar{\epsilon}$ and the generator $R_{1}^{i}(q)$ will be
replaced by $R_{2}^{i}(q)$.

\section{Supersymmetric harmonic oscillator}

\label{SUSYharm}

In this section, we analyse an $N=1$ supersymmetric free toy model with $%
(1,1)$ physical degrees of freedom described by one bosonic $x$ and two
fermionic $\psi ,\bar{\psi}$ coordinates: collectively, $q^{i}=(x,\psi ,\bar{%
\psi})$, $i=1,...,n;n=(1,2)$, from the generalized SUSY perspectives. Here,
we find that (trivial) interaction terms for such a supersymmetric model
emerge naturally through the Jacobian of functional measure. Let us start by
writing the classical action $S(q)$ for a supersymmetric harmonic oscillator:%
\begin{equation}
S=\int_{t_{in}}^{t_{out}}dt\left[ \frac{1}{2}\dot{x}^{2}-\frac{1}{2}\omega
^{2}x^{2}+i\bar{\psi}\dot{\psi}-\omega \bar{\psi}\psi \right] .
\end{equation}%
This action refers to a free toy model with $(1,1)$ physical degrees of
freedom, formally due to the presence of second-class constraints for $\psi $%
. Here, we pass to dimensionless quantities, so that, for convenience, the
mass is $m=1$ for the bosonic part. The action is invariant up to a total
time derivative with respect to an $N=1$ subalgebra of the total SUSY
superalgebra with parameter $\epsilon ^{1}=\epsilon $ ($m=1$),
\begin{equation}
\delta _{\epsilon }\big[x,\psi ,\bar{\psi}\big]\ =\ \frac{1}{\sqrt{2}}\left[
\bar{\psi},-\imath \dot{x}-\omega x,\,0\right] \epsilon \equiv \big[x,\psi ,%
\bar{\psi}\big]\overleftarrow{s}\epsilon ,  \label{susy}
\end{equation}%
and also with respect to an $N=1$ subalgebra with an odd parameter $\epsilon
^{2}=\bar{\epsilon}$,
\begin{equation}
\delta _{\bar{\epsilon}}\big[x,\psi ,\bar{\psi}\big]\ =\ -\frac{1}{\sqrt{2}}%
\left[ \psi \bar{\epsilon},\,0,\,-\imath \dot{x}+\omega x\right] \bar{%
\epsilon}\equiv \big[x,\psi ,\bar{\psi}\big]\overleftarrow{\bar{s}}\bar{%
\epsilon}\ ,  \label{susy1}
\end{equation}%
which relates with the above subalgebra by means of complex conjugation, $%
\delta _{\bar{\epsilon}}\big[x,\psi ,\bar{\psi}\big]=(\delta _{\epsilon }%
\big[x,\bar{\psi},\psi \big])^{\ast }$, determined by the rule:
\begin{equation}
(x,\psi ,\bar{\psi},\,\epsilon ,\bar{\epsilon})^{\ast }=(x,\bar{\psi},\,\psi
,\bar{\epsilon},\,\epsilon )\ \mathrm{and}\ (ab)^{\ast }=b^{\ast }a^{\ast }\
\mathrm{for}\ a,b\in \{x,\psi ,\bar{\psi},\,\epsilon ,\bar{\epsilon}\}.
\label{compconj}
\end{equation}%
$N=2$ SUSY algebraic transformations are determined by the identification $%
\epsilon ^{a}=(\epsilon ,\bar{\epsilon})$, $\psi ^{a}=(\psi ,\bar{\psi})$:
\begin{equation}
\delta _{\epsilon _{a}}\big[x,\psi ^{b}\big]=\big[x,\psi ^{b}\big]%
\overleftarrow{s}_{a}\epsilon ^{a}\equiv R_{a}^{i}(q)\epsilon ^{a},\ \mathrm{%
for}\ R_{a}^{i}=\frac{1}{\sqrt{2}}\big[\varepsilon _{ca}\psi
^{c},[(-1)^{b}\imath \dot{x}-\omega x]\delta _{a}^{b}\big],  \label{N2susy}
\end{equation}%
whereas $N=2$ FSUSY transformations form an Abelian group $\{g(\epsilon
^{a})=\exp \{\overleftarrow{s}_{a}\epsilon ^{a}\}\}$, and quadratic terms in
powers of $(\epsilon )^{2}=\epsilon _{a}\epsilon ^{a}=2\epsilon \bar{\epsilon%
}$ together with finite transformations realized on $q^{i}$ are%
\begin{equation}
\lbrack x,\psi ^{b}]\overleftarrow{s}^{a}\overleftarrow{s}_{a}(\epsilon
)^{2}=\left[ \omega x,\frac{1}{2}((-1)^{b}\imath {-}\omega )\psi ^{b}%
\right] (\epsilon )^{2}\Longrightarrow \Delta _{\epsilon ^{a}}[x,\psi
^{b}]=\delta _{\epsilon _{a}}\big[x,\psi ^{b}\big]+\frac{1}{4}[x,\psi ^{b}]%
\overleftarrow{s}^{a}\overleftarrow{s}_{a}(\epsilon )^{2},
\label{quadrterms}
\end{equation}%
then $S([x,\psi ^{b}]g(\epsilon ^{a}))=S(x,\psi ^{b})$ for arbitrary finite
FD $\epsilon ^{a}$. SUSY invariant interaction terms (with respect to
algebraic transformations) may be given by (for $n>1$) polynomial in $x,\psi
,\bar{\psi}$
\begin{eqnarray}
\hspace{-0.5em}S_{int} &\hspace{-0.7em}=&\hspace{-0.6em}\int_{t_{in}}^{t_{out}}dt\sum_{n=2}^{M}g_{n}\left[ -\frac{n}{%
\sqrt{2}}x^{n-1}\bar{\psi}\psi -\frac{i}{\sqrt{2}}x^{n}\dot{x}-\frac{1}{%
\sqrt{2}}\omega x^{n+1}\right] =\int_{t_{in}}^{t_{out}}dt\sum_{n=2}^{M}g_{n}%
\left[ \delta _{\epsilon }(x^{n}\psi )\right] \overleftarrow{\partial }%
_{\epsilon },  \notag \\
\hspace{-0.5em}&\hspace{-0.7em}=&\hspace{-0.6em}-\int_{t_{in}}^{t_{out}}dt\sum_{n=2}^{M}g_{n}\left[ \frac{\sqrt{2}}{(n+1)}%
\delta _{\epsilon }\delta _{\bar{\epsilon}}(x^{n+1})\right] \overleftarrow{%
\partial }_{\epsilon }\overleftarrow{\partial }_{\bar{\epsilon}%
}=-\int_{t_{in}}^{t_{out}}dt\sum_{n=2}^{M}g_{n}\left[ \frac{\sqrt{2}}{2(n+1)}%
(x^{n+1})\overleftarrow{s}^{a}\overleftarrow{s}_{a}\right],  \label{interex}
\end{eqnarray}%
with some coupling constants $g_{n}$ providing a correct dimension of the
action. The interaction (\ref{interex}) appears trivial, due to definition (%
\ref{nontrivial}). Then, the full action incorporating interaction, $%
S_{full}=S+S_{int}$, is invariant under $N=1$ FSUSY transformations given by
(\ref{susy}) and (\ref{susy1}), as well as under $N=2$ FSUSY transformations
(\ref{quadrterms}).

The generators of SUSY transformations (\ref{susy}), (\ref{susy1}) and (\ref%
{N2susy}) can be presented from a standard SUSY representation using the
supercommutator $[\ ,\ \}$ for equal times:
\begin{equation}
\delta _{\epsilon }q^{i}=i[q^{i},Q\}\epsilon ,\quad \delta _{\bar{\epsilon}%
}q^{i}=i[q^{i},\bar{Q}\}\bar{\epsilon},
\end{equation}%
with an explicit realization of supercharges
\begin{equation}
\big(Q,\,\bar{Q}\big)=\textstyle\frac{1}{\sqrt{2}}\Big(\overleftarrow{%
\partial }_{x}\bar{\psi}+\overleftarrow{\partial }_{\psi }[i\dot{x}+\omega
x],\,\overleftarrow{\partial }_{x}{\psi }+\overleftarrow{\partial }_{\bar{%
\psi}}[-i\dot{x}+\omega x\Big)].
\end{equation}%
which is nothing else than $s$ and $\bar{s}$, respectively, satisfying the
algebra (\ref{SUSYalg}).

\subsection{Generalized SUSY transformations and Jacobians}

\label{GenSUSYharm} Following Section~\ref{GenSUSY}, we generalize the SUSY
transformations (\ref{susy}), (\ref{susy1}) and (\ref{quadrterms}) by making
the transformation parameters finite and field-dependent:%
\begin{equation}
\Big(\delta _{\epsilon },\delta _{\bar{\epsilon}}\Big)q^{i}\ =\ q^{i}\Big(%
\overleftarrow{s}\epsilon (q),\overleftarrow{\bar{s}}\bar{\epsilon}(q)\Big)\
\ \ \mathrm{and}\ \ \ \Delta _{\epsilon ^{a}}q^{i}\ =\ q^{i}\Big(%
\overleftarrow{s}_{a}\epsilon ^{a}(q)+\frac{1}{4}\overleftarrow{s}^{a}%
\overleftarrow{s}_{a}\big({\epsilon }(q)\big)^{2}\Big)
\end{equation}%
Corresponding to $N=1$ and $N=2$ FSUSY transformations, the Jacobians of a
change of variables in the path integral (\ref{ZZ0}) in question are given
by (\ref{jacobianres2}):%
\begin{equation}
J(q(\epsilon ))=\left( 1+\epsilon \overleftarrow{s}\right) ^{-1}\ \mathrm{and%
}\ J\left( q(\Lambda \overleftarrow{s}_{a})\right) =\left( 1+\frac{1}{2}%
\Lambda \overleftarrow{s}_{a}\overleftarrow{s}^{a}\right) ^{-2},\ \mathrm{for%
}\ \epsilon _{a}(q)=\Lambda (q)\overleftarrow{s}_{a}.  \label{jacobianresf}
\end{equation}%
where $\Lambda (q)$ is an arbitrary bosonic functional. When considering the
receipe \cite{sdj}, the finite FD parameter $\epsilon (q)$ presented in
terms of an infinitesimal $\epsilon ^{\prime }(q(\kappa )$:
\begin{equation}
\epsilon (q)=\int_{0}^{1}d\kappa \epsilon ^{\prime }(q(\kappa ))\ \mathrm{and%
}\ \bar{\epsilon}(q)=\int_{0}^{1}d\kappa \bar{\epsilon}^{\prime }(q(\kappa ))
\label{inffdfin}
\end{equation}%
represent arbitrary finite FD SUSY parameters. The Jacobian of both $N=1$
SUSY transformations can be calculated using (\ref{sdjdet}).

This shows that the interactions terms (\ref{interex}) may be generated by $%
N=1$ and $N=2$ FSUSY transformations with appropriate  parameters.

\subsection{ Generating the interaction terms}

To find an explicit finite FD parameter $\epsilon (q)$ for $N=1$ SUSY
transformations which generates the trivial interaction terms (\ref{interex}%
), we consider the functional equation
\begin{equation}
Z_{0}=Z_{int}\ \mathrm{where}\ Z_{int}=\int \mathcal{D}q\exp \left\{ \frac{%
\imath }{\hbar }\big[S(q)+S_{int}(q)\big]\right\} ,  \label{funceq}
\end{equation}%
and $Z_{0}$ is determined in (\ref{ZZ0}). Making a change of variables in
the integrand of $Z_{0}$ generated by FSUSY, we obtain an equation with
accuracy up to a total functional derivative:
\begin{equation}
i\hbar \ln J(q(\epsilon ))+S_{int}(q)=0\Longleftrightarrow i\hbar \ln
(1+\epsilon (q)\overleftarrow{s})=\int_{t_{in}}^{t_{out}}dt%
\sum_{n=2}^{M}g_{n}\left[ x^{n}\psi \right] \overleftarrow{s},
\label{compeq}
\end{equation}%
which we call a \emph{compensation equation}. Because both parts of the
compensation equation are $s$-exact, we determine the unknown $\epsilon (q)$
in terms of $S_{int}$
\begin{equation}
\epsilon (q|S_{int})=\frac{\imath }{\hbar }g(y)\int_{t_{in}}^{t_{out}}dt%
\sum_{n=2}^{M}g_{n}\left[ x^{n}\psi \right] ,\ \mathrm{for}\ g(y)=\frac{%
1-\exp \{y\}}{y},\ y\equiv \frac{\imath }{\hbar }S_{int}.  \label{solcomp}
\end{equation}%
Vice-versa, considering equation (\ref{compeq}) for some unknown
interaction, we can always construct a trivial interaction $S_{int}=U(q)%
\overleftarrow{s}$ for any $N=1$ FSUSY transformation with a given $\epsilon
(q)$:
\begin{equation}
U(q|\epsilon )\ =\ \textstyle\frac{\hbar }{\imath }\left[ \sum\nolimits_{n=1}%
\textstyle\frac{(-1)^{n-1}}{n}\left( \epsilon \overleftarrow{s}\right)
{}^{n-1}\right] \epsilon .  \label{comeq}
\end{equation}%
The same can be done for an $N=1$ FSUSY with $\bar{\epsilon}(q)$ concerning
a one-to-one correspondence among trivial interactions, represented as $\bar{%
U}(q)\overleftarrow{\bar{s}}$, $\varepsilon (\bar{U}(q))=1$, and a set of
respective $N=1$ FSUSY transformations.

Concerning the case of $N=2$ FSUSY transformations with ${\epsilon }^{a}(q)$%
, $a=1,2$, the generation of trivial $N=2$ supersymmetric interactions is
the same for functionally-dependent ${\epsilon }^{a}(q)=\Lambda
\overleftarrow{s}{}^{a}$. The corresponding compensation equation to provide
(\ref{funceq}) and its solution for a given interaction (\ref{interex}) with
bosonic the potential $U_{2}(q)=\sum_{n=2}^{M}g_{n}\frac{\sqrt{2}}{2(n+1)}%
x^{n+1}$: $S_{int}=U_{2}(q)\overleftarrow{s}^{a}\overleftarrow{s}_{a}$,
\begin{eqnarray}
&&i\hbar \ln J(q(\epsilon ))+S_{int}(q)=0\Longleftrightarrow i\hbar \ln (1+%
\frac{1}{2}\Lambda (q)\overleftarrow{s}_{a}\overleftarrow{s}^{a})=-U_{2}(q)%
\overleftarrow{s}^{a}\overleftarrow{s}_{a},\ ,  \label{compeq2} \\
&&\Lambda (q|U_{2})\overleftarrow{s}_{a}=\epsilon _{a}(q|U_{2})=\frac{i}{%
2\hbar }g(y)U_{2}\overleftarrow{s}_{a},\ \Lambda (q|U_{2})=\frac{i}{2\hbar }%
g(y)U_{2},\ \mathrm{for}\ y\equiv ({i}/{4\hbar })U_{2}\overleftarrow{s}{}^{a}%
\overleftarrow{s}{}_{a}.  \label{compeqsol2}
\end{eqnarray}%
Conversely, for an unknown interaction we can always construct a trivial
interaction, $S_{int}=U_{2}(q)\overleftarrow{s}^{a}\overleftarrow{s}_{a}$,
for any $N=2$ FSUSY transformation with a given $\epsilon ^{a}(q)=\Lambda
\overleftarrow{s}^{a}$:
\begin{equation}
U_{2}(q|\epsilon _{a})=4\imath {\hbar }\left[ \sum\nolimits_{n=1}\textstyle%
\frac{(-1)^{n-1}}{2^{n}n}\left( \Lambda \overleftarrow{s}^{a}\overleftarrow{s%
}_{a}\right) {}^{n-1}\Lambda \right] .  \label{comeqsp}
\end{equation}%
Therefore, if the trivial interaction $S_{tr}$ is given by $S_{tr}=U%
\overleftarrow{s}=\bar{U}\overleftarrow{\bar{s}}=U_{2}\overleftarrow{s}{}^{a}%
\overleftarrow{s}{}_{a}$ then it can be generated (or removed from the
initial action) by any $N=1,2$ FSUSY transformation with respective $%
\epsilon (q),\bar{\epsilon}(q),\epsilon ^{a}(q)=\Lambda \overleftarrow{s}{}%
^{a}$.

Omitting the details of a similar application of $N=1$ FSUSY transformations
in the form (\ref{inffdfin}) to derive interaction (\ref{interex}), we
stress that solving the problem amounts to calculating the Jacobian $%
J(\kappa )$ in equation (\ref{sdjdet}). To find an unknown $S_{1}(q(\kappa
),\kappa )$, we consider an infinitesimal FD parameter in the form
\begin{equation}
\epsilon^{\prime} = -\int_{t_{in}}^{t_{out}}dt\sum_{n=2}^{M}(g_{n}x^{n}\psi ).
\end{equation}%
We then choose the following ansatz for $S_{1}$:
\begin{equation}
S_{1}(q(\kappa ),\kappa )= - \int_{t_{in}}^{t_{out}}dt\sum_{n=2}^{M}g_{n}\left( \chi _{1}(\kappa
)x^{n-1}\bar{\psi}\psi +\chi _{2}(\kappa )x^{n}\dot{x}+\chi _{3}(\kappa
)x^{n+1}\right) ,
\end{equation}%
where $\chi _{i},i=1,2,3$ are constant $\kappa $-dependent parameters
satisfying the condition $\chi _{i}(\kappa =0)=0$. From equation (\ref{jac}%
), we derive the following differential (in $\kappa $) equations:
\begin{equation}
\sqrt{2}\chi _{1}^{\prime }-{n}=0,\ \sqrt{2}\chi _{2}^{\prime }-{\imath }%
=0,\ \sqrt{2}\chi _{3}^{\prime }-\omega =0,
\end{equation}%
whose obvious solution (as one integrates from $0$ to $\kappa $)
\begin{equation}
\Big(\chi _{1},\,\chi _{2},\,\chi _{3}\Big)=\Big(\textstyle\frac{n}{\sqrt{2}}%
\kappa ,\,\frac{i}{\sqrt{2}}\kappa ,\,\frac{1}{\sqrt{2}}\omega \kappa \Big),
\end{equation}%
leads to an explicit form of $S_{1}(q(\kappa ),\kappa )$, while $\kappa =1$
leads to $S_{int}$ (\ref{interex}).

\section{Conclusions}

We have extended the results and ideas of our previous study \cite{sdj, re,
re1} considering special Abelian SUSY transformations as a symmetry of a
Lagrangian action with bosonic and fermionic degrees of freedom, which form
a superalgebra with $m$ Grassman-odd parameters. The SUSY invariance of the
action for infinitesimal values of the parameters is restored to the case of
finite values by solving the Lie equations. As a result, we have
constructed, starting from a Lie superalgebra, a Lie supergroup [where $\exp
$-correspondence completely maps the Lie superalgebra to the Lie supergroup (%
\ref{Liemod})] with each of its element being an invariance transformation
of the supersymmetric action in powers of the Grassman-odd parameters. This
construction generalizes the case of BRST ($m=1$) and BRST-antiBRST ($m=2$)
finite transformations for gauge theories with a closed gauge algebra,
including Yang--Mills theories. We have calculated the Jacobian of a change
of variables in the path integral with a supersymmetric action, given by
finite SUSY transformations with field-dependent parameters in (\ref%
{jacobianres}), which contains as a partial case the Jacobians of formal
BRST and BRST-antiBRST finite FD transformations. Because the set of FSUSY
transformations satisfies the equivalence theorem \cite{tyutinkallosh}
conditions, the addition from the functional measure in the path integral
may modify the supersymmetric action by a Jacobian more than quadratic in
powers of fields that still leads to the same conventional $S$-matrix. We
have called such additions to the action trivial interactions. Non-trivial
FSUSY invariant interactions cannot be generated by this receipt. It is
shown that the presence of $m$-parametric FSUSY transformations leads to the
presence of standard Ward identities for generating functionals of Green
functions (\ref{usWI}) corresponding to constant odd parameters, as well as
to modified Ward identities (\ref{modWI}) depending on FD finite odd
parameters $\epsilon^\alpha (q)$.

We have illustrated these results by a simple free toy model with $(1,1)$
physical degrees of freedom describing a supersymmetric harmonic oscillator
by a generalization of $N=1$ and $N=2$ SUSY transformations. It is shown
that any trivial interaction can be completely generated from the functional
measure by means of $N=1$ and $N=2$ FSUSY transformations respectively with
FD parameter and functionally-dependent parameters.

The present research may be used to analyse the influence on the structure
of a quantum action of real space-time SUSY transformations with FD
parameters, which, however, do not form an Abelian superalgebra and contain,
in addition to $Q$ and $\bar{Q}$, also a Grassman-even generator of momenta,
$P^{\mu }$. At the same time, in the case of additional presence of gauge
invariance for a supersymmetric action the problem of a joint consideration
of FSUSY transformations and BRST or BRSTantiBRST transformations for a
quantum action may prove to be a promicing direction of reasearch.

\vspace{1ex} \noindent \textbf{Acknowledgments \ } The authors are grateful
to P.Yu. Moshin for interest and discussions. The study of A.R. was
supported by the RFBR grant No. 16-42-700702.


\begin{thebibliography}{99}
\bibitem{superstring} M. Green, J. Schwarz and E. Witten, \textit{%
Superstring theory}, Cambridge University Press, (1987).

\bibitem{fer} S. Ferrara, D. Freedman and P. van Nieuwenhuizen, Phys. Rev. D
13, 3214 (1976).

\bibitem{des} S. Deser and B. Zumino, Phys. Lett. B 62, 335 (1976).

\bibitem{buchbinder} I. L. Buchbinder, E. A. Ivanov and I. B. Samsonov,
arXiv:1603.02768 [hep-th].

\bibitem{ivanov} E. A. Ivanov, arXiv: 1604.01379 [hep-th].

\bibitem{reviewsV} M. Vasiliev, Lect. Notes Phys. 892, 227 (2015).

\bibitem{reviews3} A. Fotopoulos and M. Tsulaia, Int. J. Mod. Phys. A. 24, 1
(2008), arXiv:0805.1346[hep-th].

\bibitem{symferm-ads} I. L. Buchbinder, V. A.Krykhtin and A.A.Reshetnyak,
Nucl.Phys. B. 787, 211 (2007).

\bibitem{Reshetnyk2} A. A. Reshetnyak, Nucl. Phys. B. 869, 523 (2013).

\bibitem{MetsaevBV1} R.R. Metsaev, Phys.Lett. B. 720, 237 (2013).

\bibitem{Zinoviev} I. L. Buchbinder, T.V. Snegirev and Yu.M. Zinoviev, JHEP.
1510, 148 (2015).

\bibitem{avi} F. Cooper, A. Khare, and U. P. Sukhatme, \textit{Supersymmetry
in quantum mechanics}, World Scientific, Singapore (2001).

\bibitem{wit} E. Witten, Nucl. Phys. B 188, 513 (1981).

\bibitem{coo} F. Cooper and B. Freedman, Annls. Phys. 146, 262 (1983).

\bibitem{sal} P. Salomonson and J. van Holten, Nucl. Phys. B 196, 509 (1982).

\bibitem{31} M. Bernstein and L. Brown, Phys. Rev. Lett. 52, 1933 (1984).

\bibitem{32} P. Kumar, M. Ruiz-Altaba and B.S. Thomas, Phys. Rev. Lett. 57,
2749 (1986).

\bibitem{33} F. Marchesoni, P. Sodano and M. Zannetti, Phys. Rev. Lett. 61,
1143 (1988).

\bibitem{34} W.-Y. Keung, E. Kovacs and U. Sukhatme, Phys. Rev. Lett. 60, 41
(1988).

\bibitem{brs} C. Becchi, A. Rouet and R. Stora, Phys. Lett. B 52, 344
(1974); Annals Phys. {98}, 287 (1976).

\bibitem{brs1} I. V. Tyutin, Lebedev Inst. preprint. 39 (1975), arXiv:
0812.0580 [hep-th].

\bibitem{rup} C. Rupp, R. Scharf and K. Sibold, Nucl. Phys. B 612, 313
(2001).

\bibitem{van} P. Van Nieuwenhuizen, arXiv: hep-th/0408179.

\bibitem{deWittH} de Witt, B., van Holten, J.W.. Phys. Lett. B79, 389 (1979).

\bibitem{bv} I.A. Batalin and G.A.Vilkovisky, Phys. Lett. B 102, 27 (1981).

\bibitem{bfv} E.S. Fradkin and G.A. Vilkovisky, Phys. Lett. {B55, } 224
(1975)

\bibitem{bfv1} I.A. Batalin and G.A.Vilkovisky, Phys. Lett. B 69, 309 (1977).

\bibitem{ht} M. Henneaux and C. Teitelboim, \textit{Quantization of gauge
systems}, Princeton, USA: Univ. Press (1992).

\bibitem{wei} S. Weinberg, \textit{\ The quantum theory of fields, Vol-II:
Modern applications}, Cambridge, UK Univ. Press (1996).

\bibitem{sdj} S. D. Joglekar and B. P. Mandal, {Phys. Rev.} {D 51}, 1919
(1995).

\bibitem{upa} S. Upadhyay, Phys. Lett. B 723, 470 (2013); Annls. Phys. 344,
290 (2014).

\bibitem{sdj1} S. D. Joglekar and B. P. Mandal, Int. J. Mod. Phys. A 17,
1279 (2002).

\bibitem{rb} R. Banerjee and B. P. Mandal, Phys. Lett. B 488, 27 (2000).

\bibitem{susk} S. Upadhyay, S. K. Rai and B. P. Mandal, J. Math. Phys. {52},
{022301} (2011).

\bibitem{sb} S. Upadhyay and B. P. Mandal, Phys. Lett. B 744, 231 (2015);
arXiv:1407.2017 [hep-th]; Prog. Theor. Exp. Phys. 053B04 (2014); Eur. Phys.
J. {C 72}, 2065 (2012); Annals of Physics {327}, 2885 (2012); EPL {93}, {%
31001} (2011); Mod. Phys. Lett. {A 25}, {3347} (2010).

\bibitem{bss} B. P. Mandal, S. K. Rai, and S. Upadhyay, EPL 92, 21001 (2010).

\bibitem{smm} S. Upadhyay, M. K. Dwivedi and B. P. Mandal, Int. J. Mod.
Phys. A 28, 1350033 (2013).

\bibitem{fs} M. Faizal, B. P. Mandal and S. Upadhyay, Phys. Lett. B 721, 159
(2013).

\bibitem{sud1} R. Banerjee, B. Paul and S. Upadhyay, Phys. Rev. D 88, 065
019 (2013).

\bibitem{sudhak} S. Upadhyay, Annls. Phys. 356, 299 (2015); Phys. Lett. B
740, 341 (2015); EPL 105, 21001 (2014); Phys. Lett. B 727, 293 (2013); EPL
104, 61001 (2013); arXiv:1308.0982 [hep-th].

\bibitem{rbs} R. Banerjee, B. Paul, S. Upadhyay, Phys. Rev. D 88 (2013)
065019.

\bibitem{das} S. Upadhyay and D. Das, Phys. Lett. B 733, 63 (2014).

\bibitem{rs} R. Banerjee and S. Upadhyay, Phys. Lett. B 734, 369 (2014).

\bibitem{su} S. Upadhyay, Annals. Phys. 340, 110 (2014).

\bibitem{llr1} P. Lavrov, O. Lechtenfeld, A. Reshetnyak, JHEP. 1110, 043
(2011).

\bibitem{Gribov} V.N. Gribov, Nucl. Phys. B. 139, 1 (1978).

\bibitem{Zwanziger} D. Zwanziger, Nucl. Phys. B. 323, 513 (1989).

\bibitem{ll1} P. Lavrov and O. Lechtenfeld, Phys. Lett. B. 725, 382 (2013).

\bibitem{re} A. Reshetnyak, Int. J. Mod. Phys. A 29, 1450128 (2014).

\bibitem{aBRST1} G. Curci and R. Ferrari, Phys. Lett. B. 63, 91 (1976).

\bibitem{aBRST2} L. Alvarez-Gaume and L. Baulieu, Nucl. Phys. B. 212, 255
(1983).

\bibitem{aBRST3} V.P. Spiridonov, Nucl. Phys. B. 308, 527 (1988).

\bibitem{re1} P. Yu. Moshin and A. A. Reshetnyak, Nucl. Phys. B. 888, 92
(2014); Phys. Lett. B 739, 110 (2014); Int. J. Mod. Phys. A 29, 1450159
(2014); Int. J. Mod. Phys. A 30, 1550021 (2014).

\bibitem{re5} P. Yu. Moshin and A. A. Reshetnyak, arXiv: 1506.04660 [hep-th].

\bibitem{mr} P. Yu. Moshin and A. A. Reshetnyak, arXiv: 1604.03027 [hep-th].

\bibitem{fp} L.D. Faddeev and V.N. Popov, Phys. Lett. B. 25, 29 (1967).

\bibitem{blt} I.A. Batalin, P.M. Lavrov and I.V. Tyutin, Int. J. Mod. Phys.
A. 29, 1450166 (2014).

\bibitem{tyutinkallosh} K.E. Kallosh and I.V. Tyutin, Sov. J. Nucl. Phys.
17, 98 (1973).
\end{thebibliography}
\end{document}